# IEEE 802.1AS Clock Synchronization Performance Evaluation of an Integrated Wired-Wireless TSN Architecture

Iñaki Val, *Senior Member, IEEE*, Óscar Seijo, Raul Torrego and Armando Astarloa

*Abstract*— **Industrial control systems present numerous challenges from the communication systems perspective: clock synchronization, deterministic behavior, low latency, high reliability, flexibility, and scalability. These challenges are mostly solved with standard technologies over Ethernet, e.g., Time-Sensitive Networking (TSN). As a research trend, it is expected that TSN will converge with wireless, leading to the Wireless TSN paradigm. Also, Wireless TSN is expected to be integrated with Ethernet TSN to create large-scale wired-wireless (Hybrid) TSN networks. The first step towards Hybrid TSN is the distribution of the clock reference from the wired to the wireless domain. In this paper, we leverage existing Ethernet TSN and wireless technologies implementations (Wi-Fi and w-SHARP) and we present two hardware architectures specifically engineered to enable the clock synchronization distribution among the network domains. The hardware architectures have been implemented over a System-on-Chip (SoC) Field Programmable Gate Array (FPGA) platform. We demonstrate through several experiments that the implementation is able to fulfill the synchronization performance required by TSN.**

## I. Introduction

IEEE 802.1 Time-Sensitive Networking (TSN) is dramatically changing the way industrial networks and applications are designed and implemented [1]. TSN enables the coexistence of time-critical and best-effort traffic over Ethernet (Ethernet TSN) while keeping the strict behavior required by time-critical applications. TSN comprises more than 20 standards focused on different functionalities. Among others: synchronization (802.1AS-2020 [2]), traffic shaping and scheduling (802.1Qbv [3]), and network management (802.1Qcc [4]).

In parallel to the TSN standardization process, the industry and academia are following a new research trend towards the development and adoption of wireless technologies with TSN-like capabilities, or wireless TSN [5]. Currently, 5G and 802.11 are considered the main candidates to enable wireless TSN [6]. In addition to that, both 5G and 802.11 are expected to be integrated with Ethernet TSN towards a Hybrid TSN technology [7]. In essence, the ultimate objective of wireless TSN research is to enable support for TSN functionalities across a wide range of wireless protocols. Hybrid TSN presents significant advantages, such as enhanced flexibility, low commissioning costs, and seamless interoperability of different devices, no matter whether they use a wired or wireless interface [6].

Two integration models are currently being considered in Hybrid TSN, as shown in Fig. 1: wireless TSN at the edge or as a bridge. In the edge model, the last mile connection of the network is offered by Wireless TSN, whereas in the bridge model, the Wireless TSN interconnects two separate Ethernet TSN networks. Both topologies share the TSN configuration entities, namely the Centralized User Configuration (CUC) and Centralized Network Configuration (CNC). These entities are responsible for analyzing the traffic requirements and network capabilities to generate an appropriate TSN scheduler configuration according to the users' needs.

Typically, the clock synchronization is considered as the TSN core functionality since TSN is built upon the condition of a common time shared among the network [8]. Thus, the first step to enable Hybrid TSN is to support a precise synchronization in both the wired and wireless networks and a seamless translation of the clock synchronization between them. However, the stringent synchronization requirement for TSN in industrial use cases, which ranges from 0.1 to 1 μs of maximum time deviation between the Grand Master Clock (GMC) and the slaves [9], is a real challenge in Hybrid TSN networks.

Despite that Hybrid TSN may be the future of industrial communications, very few works analyze the issues of clock synchronization translation between Ethernet TSN and the dominant wireless technologies, 802.11 and 5G [10]. In addition to that, none of them presents real implementations and measurements of the clock synchronization translation among the wired-wireless domains.

Manuscript received December, 2020; revised April, 2021; accepted August 9, 2021. Date of publication xxxx xx, 202x; date of current version August, 2021. Paper no. TII-20-5626. The work of Iñaki Val, Óscar Seijo and Raul Torrego was supported by B-INDUSTRY5G (ELKARTEK) projects of the Basque Government (Spain). The authors would like to thank the anonymous reviewers and editors for their valuable comments that have resulted in an improvement of the clarity of this paper.

I. Val, Ó. Seijo and R. Torrego are with Ikerlan Technology Research Centre, Basque Research and Technology Alliance (BRTA), P.º J.M. Arizmendiarrieta, 2. 20500, Mondragón, Spain (e-mail: ival@ikerlan.es, oseijo@ikerlan.es, rtorrego@ikerlan.es).

A. Astarloa is with the Escuela de Ingenieria de Bilbao, University of the Basque Country (UPV/EHU), 48013 Bilbao, Spain (e-mail: armando.astarloa@ehu.eus).

Corresponding Author: I. Val (e-mail: ival@ikerlan.es)

Digital Object Identifier xx.xxxx/TII.xxxx.xxxxxxx

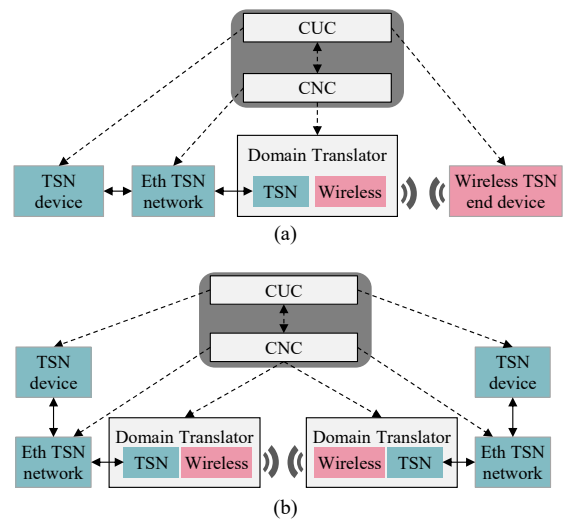

Fig. 1. Integration of Ethernet and wireless TSN at the edge (a) or as a bridge (b).

This paper presents for the first time the design and implementation of two Hardware/Software (HW/SW) network architectures with wired-wireless support that enable high-performance clock synchronization translation among the network. To do such implementation, we have leveraged current existing IPs for Xilinx System On Chip (SoC) Field Programmable Gate Arrays (FPGAs): a multiport TSN (MTSN) switch IP [11], an IEEE 802.11 modem IP [12], and a w-SHARP modem IP [13]. The first architecture comprises the MTSN switch and the 802.11 modem, while the second architecture comprises the MTSN switch and the w-SHARP modem. Over these two architectures, we have engineered the clock synchronization translation between the wired and wireless domains. We have built two HW testbeds, and we have evaluated the synchronization over different conditions. The results of these experiments show that the attainable End-to-End (E2E) clock synchronization is enough to fulfill the synchronization required by TSN and by a wide range of industrial applications.

The rest of the article is organized as follows. First, section II presents the state-of-the-art in clock synchronization for Hybrid TSN. In section III, the main techniques used to perform wireless clock synchronization are stated. The two network architectures and their implementation are shown in section IV. Section V analyzes the theoretical attainable clock synchronization performance in multi-hop hybrid networks. In section VI, the measurement setups are described. The clock synchronization results are shown in section VII. Finally, section VIII shows the conclusions of the work.

## II. BACKGROUND AND RELATED WORK

Through this section, we present in detail the clock synchronization requirements for TSN operation, and we review the state-of-the-art about clock synchronization in wireless networks and its integration with wired clock synchronization.

### A. Clock synchronization requirements

TABLE I. CLOCK SYNCHRONIZATION REQUIREMENTS OF DIFFERENT USE CASES.

| Use Case | Sync. Accuracy | E2E Latency | Devices | Area |
|---|---|---|---|---|
| Factory Automation | < 1 µs | < 1 ms | < 30 | 100 m² |
| Smart Grid | < 1 µs to 20 µs | < 10 ms to 100 ms | < 100 | 20 km² |
| Vehicular Communication | < 10 µs | N.A. | < 300 | N.A. |
| Live A/V | < 1 µs | N.A. | < 150 | 100 m² |

The TSN clock synchronization requirement is derived from the network scheduling requirements and the applications built on top of it. Typically, the most stringent synchronization comes from the network to provide guaranteed minimum latency in the time-critical traffic. For instance, IEC/IEEE 60802 [9] sets the maximum tolerable deviation from the grandmaster time to 0.1 - 1 µs for appropriate network operation in factory automation. On the application side, Table I summarizes the clock synchronization requirements for typical use cases and their characteristic parameters such as the latency, number of devices in the network, and the area of coverage [14]. The synchronization precision required by industrial applications ranges from < 1 µs for closed-loop motion control to < 20 µs for smart grid fault detection and protection between two different points of the power transmission line. Factory automation use case covers applications with isochronous real-time communication requirements, including collaborative robotics and closed-loop control. Smart grid has different application groups with different grades of requirements, where we can find fault protection, control and optimization, and monitoring applications. Finally, other relevant use cases are vehicular communications, especially for safety-related applications, and audio/video professional production equipment.

### B. Wireless clock synchronization

Several approaches have been explored by different researchers during the last few years to perform wireless clock synchronization. These approaches are commonly classified into SW-based (SW timestamps and SW clocks), or HW-based (HW timestamps and hardware clocks). SW-based approaches provide a synchronization precision in the µs order [15], whereas HW-based ones provide precisions in the range of 1 to 20 ns [16]. Unfortunately, Commercial-Off-The-Shelf (COTS) wireless cards do not usually include accessible HW timestamps and clocks.

SW-based clock synchronization over COTS wireless cards has been thoroughly explored in the last decade. For instance, [17] presents a wireless sensor network with clock synchronization capabilities. The system uses a HW timer with a resolution of 1 µs to take SW timestamps in the interrupt service routine and provides an E2E time synchronization error of 2 µs. On the 802.11 side, Mahmood et al. combined SW timestamps and Precision Time Protocol (PTP) over 802.11 to achieve a time synchronization performance in the order of 0.5 µs [15]. However, the experiments show a deterioration of the clock synchronization under high CPU or network loads.

As evidenced by these works, the attainable performance using SW-based techniques is not enough to fulfill the TSN clock synchronization requirement which ranges between 100 ns to 1 µs of maximum deviation [9]. In consequence, there exist active efforts from the researchers, standardization committees, and the industry towards the development of wireless cards with HW timestamping and PTP Hardware Clock (PHC). For instance, the 802.11 Fine Timing Measurement (FTM) was first introduced in the 802.11-2016 standard revision. [18] presents a comprehensive evaluation of the FTM for ranging purposes over a COTS 802.11 device. The results show that FTM can achieve precisions in the range of 1-5 meters over Line-of-Sight conditions and with low multipath propagation. However, the wireless card used in [18] does not include a PHC, which is required to provide enough clock synchronization stability.

Due to current generation of COTS wireless modems have limitations to perform wireless synchronization, custom solutions based on SoC-FPGA have stood as an alternative. In the research line of wireless synchronization over SoC-FPGA, [16] present an 802.11b modem built over a SoC-FPGA platform that includes HW timestamping with subsample precision. This design resulted in a synchronization accuracy of around one ns for static conditions. However, according to the experiments performed in [16], time-variant channels highly deteriorate its performance. As an example of an innovative solution, [19] presents an 802.11 modem combined with a low-cost FPGA. The wireless card was responsible for the whole communication, whereas the FPGA included the PHC and a module to perform HW timestamping using signals generated by the 802.11 modem. The performance of this solution is in the 500 ns range, far from the results obtained with SoC-FPGA implementations.

### C. Integration of wired and wireless clock synchronization

The integration of Ethernet and Wireless TSN was brought up a few years ago. Despite that is a highly relevant matter of study, few works are available that analyze or that provide real-world implementations of the integration of wired and wireless clock synchronization for both 5G and 802.11. On the 5G side, the clock synchronization support is standardized in the 3rd Generation Partnership Project (3GPP) TS 23.501 Release 16 [20]. In order to provide TSN E2E synchronization, the 5G System (5GS) is seen as a network bridge (Fig.1 (b)), which timestamps the ingress and egress synchronization frames with its internal GMC and calculates their residence time. This behavior is similar to the one found in a PTP transparent clock switch. To this end, there is a need for timing service over the Radio Access Network (RAN). The 5G internal system clock can be made available to User Equipment (UE) with the signaling of time information related to the absolute timing of radio frames as described in 3GPP TS 38.331 [21]. The 5GS can use a 5G internal clock, or GMC, as a time reference that is distributed over the transport network utilizing PTP Telecom Profile. Concerning the clock synchronization performance, 3GPP targets a synchronization error below 1 µs. Some authors have analyzed by simulation means the E2E TSN clock synchronization over 5G [22]. This work claims that precisions in the order of 1 to 4 µs can be obtained depending on the 5G RAN subcarrier spacing configuration.

Regarding the TSN integration over 802.11, 802.1AS-2020 [2] has standardized the use of the Fine Time Measurements (FTM) scheme to perform precise clock synchronization over 802.11. However, there are no works that analyze or demonstrate the integration of the clock synchronization between Ethernet TSN and 802.11. Finally, [23] presents a Hybrid network with deterministic E2E capabilities based on Ethernet TSN and a proprietary narrowband technology on the wireless segment. The clock synchronization is extended from the wired to the wireless segment using 1 Pulse Per Second (PPS) signal. In this case, the E2E clock synchronization is in the order of 2 µs.

As shown in the review of the state-of-the-art, wireless clock synchronization technologies with HW-based timestamping and PHC might support the TSN requirements, which are in the 0.1-1 µs order [9]. However, there are very few works that address the integration of the clock synchronization in Ethernet and Wireless TSN and none of them presents a real implementation of a highly integrated network with seamless clock synchronization translation among the domains. Therefore, this work contributes to the state-of-the-art by presenting a flexible SoC-FPGA architecture that enables the integration of Ethernet and wireless interfaces into the same SoC-FPGA, the development of a clock synchronization protocol on top of the devices of the network, and a comprehensive performance evaluation over different wireless conditions and topologies.

## III. WIRELESS CLOCK SYNCHRONIZATION

Most synchronization protocols are based on three processes: the protocol performs a messaging exchange that carries timing information (timestamps), the messages and their timestamps are used to evaluate the error between the master and slave nodes, and finally, the slaves correct their internal time to match the time of the master. Despite their simplicity, the attainable clock synchronization performance largely depends on the protocol implementation, ranging from several hundreds of microseconds to few nanoseconds [24]. This section provides an overview of the different techniques used to perform these processes and presents the particularities of clock synchronization over wireless.

### A. Synchronization schemes

Currently, there exist three main messaging schemes used to perform wireless clock synchronization over 802.11: the 802.1AS messaging as defined over Ethernet, the 802.11 FTM, and the beacon-based messaging.

*1) 802.1AS messaging*

IEEE 1588 standard, commonly known as PTP, is the de-facto standard to perform time synchronization in industrial networks over fiber or Ethernet because of its simplicity and precision in the tens of nanoseconds range using appropriate implementations. PTP provides several profiles with slightly different functionalities for different applications, and 802.1AS-2020 [2] is the standard that has been adopted for TSN operation.

The 802.1AS messaging exchange is shown in Fig. 2 (a). It starts with a Sync frame transmitted by the master, which takes the $t_1$ timestamp. The frame reaches the slave after traversing through the channel with a delay $t_{ms}$ and the slave takes the timestamp $t_2'$. In the case that the HW supports one-step operation, $t_1$ is directly delivered by the Sync frame. Otherwise, a Follow-up frame is sent from the master to the slave to transmit the timestamp $t_1$ to the slave (two-step operation). Afterward, the slave transmits a Delay_Req frame to obtain two more timestamps ($t_3', t_4$). The channel delay between the slave and the master is $t_{sm}$. Finally, the Delay_Resp delivers the $t_4$ timestamp to the slave.

*2) 802.11 FTM messaging*

The main purpose of this messaging protocol was to perform ranging measurements based on radio frequency localization between 802.11 devices. This protocol can be also used to perform clock synchronization and was adopted by 802.1AS over 802.11 [8]. The protocol follows a messaging scheme similar to the 802.1AS messaging, though exploiting the intrinsic 802.11 frame exchange based on data + ACK frames (Fig. 2 (b)). FTM also defines the FTM bursts, which allows sending several FTM messages in a row, reducing the error of the timing calculation by averaging several timestamps.

*3) Beacon-based messaging*

Finally, an alternative synchronization scheme with minimum radio resource consumption is based on the broadcasting of beacon frames and without feedback information from the slaves to compute the channel delay (see Fig. 2 (c)). In this scheme, the master node periodically ($T_{sync}$) generates a beacon frame that contains an egress timestamp ($t_1$). The frame reaches the slave $t_{ms}$ later and the slave takes the timestamp $t_2'$. The beacon-based messaging is one-way, i.e., the slave does not answer the synchronization frames from the master. Consequently, and as opposed to 802.1AS and 802.11 FTM messaging schemes, the devices using this scheme are not able to estimate the channel delay. This scheme is widely considered in wireless networks located at the edge, due to its low resource consumption and implementation simplicity, but it is not usually suitable for long-range communications or applications with very stringent clock synchronization requirements.

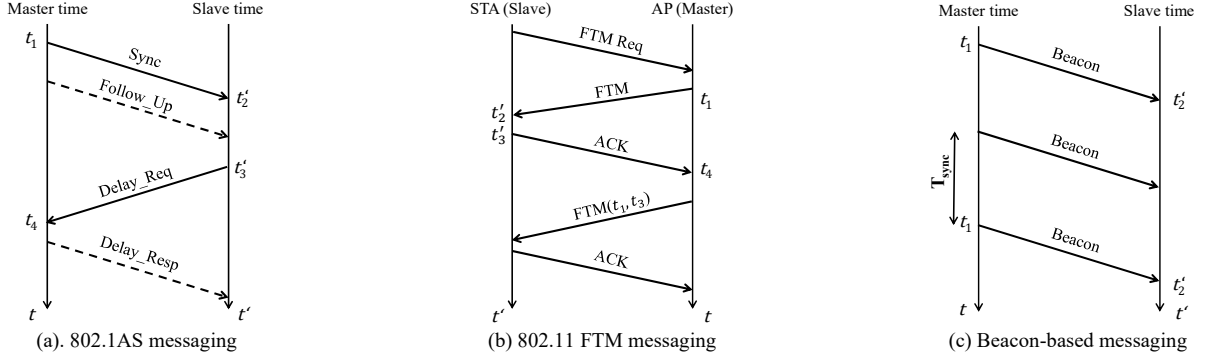

Fig. 2. Messaging schemes used in different clock synchronization protocols, being $t$ the time of the master clock and $t'$ the time of the slave clock.

### B. Error estimation and correction

Every time that a frame exchange is successfully performed, the slave node must evaluate its clock synchronization error and adjust its clock to match the desired clock value. To do so, the slave performs the next calculations

$$\tilde{t}_{ms} = \frac{t'_2 - t_1 + t_4 - t'_3}{2}, \quad (1)$$

$$\tilde{t}_o = t'_2 - t_1 - \tilde{t}_{ms}, \quad (2)$$

where $\tilde{t}_{ms}$ represents the estimated channel delay between the master and the slave and $\tilde{t}_o$ is the estimated error between the slave time and the master time. In the case of the Beacon-based messaging $\tilde{t}_{ms}$ cannot be estimated and it is either pre-calibrated or considered negligible. $\tilde{t}_o$ is commonly filtered to enhance the synchronization results. The most common filters are based on Proportional Integral (PI) filtering or linear regression. Lastly, the filtered estimation is used to adjust the clock time. The error correction process is carried out every time that a frame exchange is successfully performed to track the imperfections in the HW, such as the drift variation of the oscillators.

### C. Clock synchronization particularities over wireless

As opposed to wired, wireless propagation presents several particularities that have a direct impact on the attainable timestamping precision and thus in the clock synchronization performance [12]. Since the wireless propagation only affects the ingress timestamping process, HW-based egress and ingress timestamps have largely different precisions. Basically, egress timestamps have very high precision because the transmitter can be designed to precisely know the exact instant of transmission of the synchronization frames. Typically, egress timestamps precision is not affected by the system BW and instead only depends on 1) calibration bias (difference between the actual time of departure of the frame and the timestamp value) and 2) to digital to analog conversion jitter. The jitter is in the range of tens of picoseconds and can be considered negligible [25], whereas the calibration bias can range from below 1 ns to tens of nanoseconds. The calibration procedure is out of the scope of this paper, though several works that address the calibration in PTP over Ethernet can be followed for calibration over wireless [26].

On the contrary, ingress timestamps are not only affected by jitter and calibration bias, but also by the wireless propagation conditions and by the receiver resolution bound. Consequently, ingress timestamps are the main bottleneck in the attainable clock synchronization. The main error sources in ingress timestamping are discussed as follows.

- **Timestamping resolution bound**. Wireless digital baseband processors operate in a sample-by-sample fashion and the preamble detector, used to determine the start of a frame, is also designed in that way. Due to this, HW timestamping implementations based on preamble detection present a resolution bound equal to the baseband processor sampling period [25]. Typically, the sampling period ($T_s$) is a multiple of the inverse of the system BW. For instance, the 802.11g modem IP used in this work has 20 MHz BW and a sampling period of 50 ns.

- **Multipath propagation**. Wireless indoor propagation is typically characterized to be time-dispersive because of the different reflective elements of the environment. As a result, multiple copies of the transmitted signals affected by different delays and attenuations are captured by the receiver antenna. These reflections can introduce a significant jitter in the frame detector, especially under Non-Line-of-Sight conditions.

- **Wireless channel variation over time**. The movement of the wireless nodes or the environment induces variations in the channel impulse response. The dynamic of the channel is characterized by the coherence time $T_c$, and it defines the time duration over which the channel is considered to be invariant. This channel variation can introduce an error in the channel delay estimation and, consequently, in the clock synchronization.

## IV. SYSTEM ARCHITECTURE AND IMPLEMENTATION

We have implemented two architectures to evaluate the clock synchronization performance in Hybrid TSN networks. The first one comprises an MTSN switch IP [11] and a 802.11g modem IP [12], and the second one is a SHARP architecture [27] and comprises the same MTSN switch IP and a w-SHARP modem IP. Both architectures have been implemented over the ADRV9361-Z7035 platform completed with an Ethernet FPGA Mezzanine Card (FMC) by Opsero. The ADRV9361-Z7035 platform features a Xilinx Zynq System-on-Chip (SoC) and an Analog Devices AD9361 radio chip. The Zynq SoC comprises programmable logic (FPGA) and a dual-core ARM A9 microcontroller. The AD9361 radio chip provides 56 MHz BW and a configurable carrier frequency from 70 MHz to 6 GHz.

Subsection IV.A presents the main implementation blocks of the two architectures: the 802.11g modem, the w-SHARP modem, and the MTSN switch IP are briefly described.

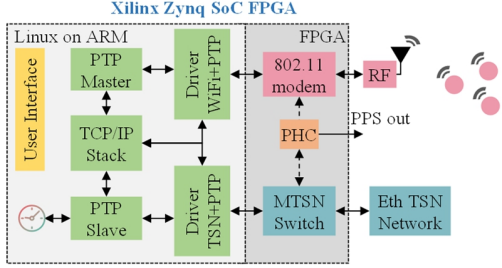

Fig. 3. Domain translator between Ethernet TSN and 802.11 (Access Point).

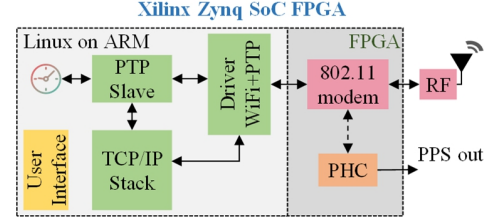

Fig. 4. 802.11 End device (STA) of the TSN – IEEE 802.11 architecture.

Afterward, Subsections IV.B and IV.C detail the integration of the main blocks of two clock-synchronized hybrid networks.

A. Main building blocks

- **MTSN Switch IP**

MTSN Switch IP core is a flexible HDL code ready to generate TSN end-point or up-to 32x port bridge implementations [11]. MTSN Switch IP supports key TSN features like IEEE 802.1AS-2020 for clock synchronization IEEE 802.1Qav for Reserved Traffic, IEEE 802.1Qbv for Scheduled Traffic, and a wide number of optional features like IEEE 802.1Qcc for Network Management, IEEE 802.1Qci for Stream Filtering and Policing, IEEE 802.1s for Multiple Spanning Tree Protocol combined with IEEE 802.1CB for Frame Replication and Elimination for Reliability and IEEE 802.1Qbu/802.3br for Frame Preemption. The current implementation features a HW timestamping unit with 8 ns of resolution and a PHC with 1 nanosecond resolution, clocked at 125 MHz.

- **IEEE 802.11g modem IP**

The 802.11g modem IP is fully compliant with the IEEE 802.11g standard specification. The PHY provides 20 MHz BW, supports data rates from 6 Mbps (Modulation and Coding Scheme (MCS) 0) up to 54 Mbps (MCS 7), and provides a demodulation latency below 16 μs. The MAC implements the whole Carrier Sense Multiple Access with Collision Avoidance (CSMA/CA) process as defined by IEEE 802.11 standard, including random back-off time, Network Allocation Vector (NAV), ACK transmission, and retransmissions. The IEEE 802.11g modem uses the 802.1AS messaging protocol configured in unicast mode to perform the wireless clock synchronization between the Access Point (AP), and the Stations (STAs). The 802.11g modem IP features a HW timestamping unit with a resolution of 50 ns and a PHC with 1 nanosecond resolution, clocked at 160 MHz. The current version of the modem is limited to 20 MHz because it is based on the 802.11g standard. Newer 802.11 releases (e.g., 802.11ax) which provide up to 160 MHz of BW, could enhance the timestamping resolution from 50 ns to 6.25 ns (1/160 MHz).

- **w-SHARP modem IP**

w-SHARP is a high-performance Wireless TSN solution specifically optimized for industrial applications that are characterized by the generation of small payloads and time-critical requirements. w-SHARP provides TSN-like capabilities such as deterministic and scheduled transmissions, high reliability, and precise synchronization [27]. The w-SHARP modem IP provides 20 MHz BW, cycle times from several milliseconds up to 100 μs, transmission of frames as short as 8 μs, E2E latency up to 50 μs, and a PER below $10^{-6}$ in realistic industrial environments without interferences. w-SHARP uses a beacon-based messaging protocol combined with HW timestamping and PHC to perform the synchronization between the master and the slaves. The w-SHARP modem IP HW timestamping unit has a resolution of 50 ns, whereas the PHC has a numerical resolution of 1 ns, clocked at 160 MHz. The latest release of the w-SHARP IP requires a real-time Operating System (OS) working on top of the IP. Nonetheless, it is expected that the next release will operate only with Linux.

B. Ethernet TSN – IEEE 802.11 architecture

The Ethernet TSN – IEEE 802.11 architecture enables the integration of Ethernet TSN and 802.11 standard interfaces in the same network. The main element of the architecture is the clock domain translator, which provides the link between the wired and wireless domains (see Fig. 3). Its main functionalities are the global time transference from the Ethernet TSN to the 802.11 network and the routing of best-effort data. Due to 802.11g lacks determinism, this architecture cannot be considered a Hybrid TSN solution. The incoming/outgoing Ethernet TSN traffic is routed through Linux TCP/IP stack, as regular traffic, losing hard real-time capabilities. Even so, this configuration can be used to evaluate the clock synchronization performance in a hybrid network.

The Ethernet TSN – IEEE 802.11 domain translator is configured as a boundary clock. On the one hand, the Ethernet TSN interface acts as a PTP slave, i.e. it receives the timing information from a GMC connected to the TSN interface and synchronizes the local PHC. On the other hand, the 802.11 interface is a PTP master, i.e., it transmits the timing information to the 802.11 network using the local PHC. The timing translation between 802.11 and Ethernet TSN is performed through the PHC of the architecture, which is shared by both 802.11 and Ethernet TSN. 802.11 only uses the PHC for timestamping, whereas Ethernet TSN uses the PHC for timestamping and synchronization.

Regarding the HW design, the MTSN and 802.11 modem IPs are driven by two different digital clock sources ($clk_{MTSN}$ and $clk_{Wireless}$) at 125 and 160 MHz respectively. The PHC is driven by $clk_{MTSN}$ and therefore the propagation of its time counter to the 802.11 modem is not straightforward. Special attention must be taken to deliver the PHC time counter output to the 802.11 modem. The complete design is driven by the same oscillator source of 40 MHz, but without loss of generality, we will assume that the clk$_{MTSN}$ and clk$_{Wireless}$ are asynchronous to each other. In order to read the PHC timer counter value from a different digital clock domain, a clock domain crossing (CDC) circuits is required to avoid metastability issues [28]. Fig. 5 shows the synchronizer circuit used in the current design, which reduces the occurrence probability of a metastable situation. The synchronizer circuit maintains the data from the source stable during several destination clock periods, which ensures the correct sampling of the source data. In this particular case, the circuit enables the reading of the TSN PHC from the 802.11 modem side. This circuit requires that the clock source frequency is at least a fourth part of the destination frequency. Therefore, the origin PHC time, which is locally generated at 125 MHz, is downsampled to 31.25 MHz in the CDC circuit.

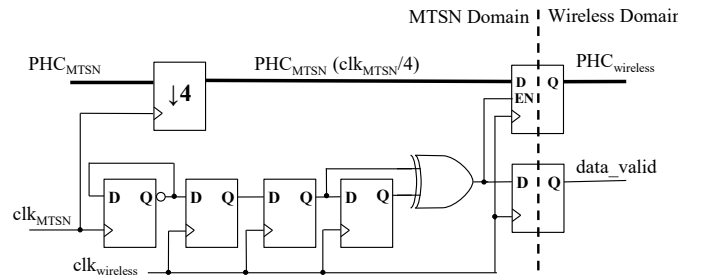

Fig. 5. Clock Domain Crossing (CDC) circuit.

The timestamping process is done directly in HW by the IPs, and these timestamps are transferred to two PTP daemons using the Linux Kernel network interface. Besides, the PHC time operations are commanded through the Linux Kernel PTP interface. These operations are used for getting/setting the time and the frequency drift of the PHC.

Regarding the SW, the clock domain translator has two PTP daemon instances. The first one is configured as a slave (Ethernet TSN), and the second one as master (802.11). The Ethernet TSN daemon has two tasks: first, it evaluates the clock synchronization error based on the timestamps generated on the Ethernet TSN interface and, second, it corrects the PHC offset and frequency drift. Once the PHC is synchronized to the Ethernet TSN network, the 802.11 PTP daemon uses the PTP and network interface from the 802.11 Linux IEEE80211 device driver, which has been modified in order to support PTP. Both daemons use the 802.1AS messaging configured as one-step. The 802.11 PTP daemon has been configured in unicast mode due to 802.11 does not support multicast traffic. Note that we have used the Ethernet 802.1AS messaging instead of the FTM 802.11 messaging, taking into account that we want to use Linux PTP [29] software implementation without any modification. It is also worth mentioning that both FTM and 802.1AS messaging have similar performance and that FTM could also be enabled in this architecture.

The 802.11 AP provides connectivity and clock synchronization to 802.11 STAs. As shown in Fig. 4, the node architecture in the STA case is simplified compared to the AP, taking into account that it does not have the Ethernet TSN interface. In the STA, the 802.11 interface is used as a slave, i.e., it receives the timing information from the 802.11 AP. The timing information is then used to synchronize its local PHC through the Linux PTP interface. Additionally, the Linux system clock is synchronized to the PHC and any HW peripheral can be synchronized as well.

C. Ethernet TSN – w-SHARP architecture

The Ethernet TSN – w-SHARP architecture is similar to the one presented in the previous subsection. The domain translator (see Fig. 6) features an Ethernet TSN interface and a w-SHARP interface that share the same PHC. On the one hand, the Ethernet TSN synchronization matches the one used in the Ethernet TSN – IEEE 802.11 architecture. The network interface is linked to a slave PTP daemon which synchronizes the PHC through the Linux PTP interface provided by the Ethernet TSN device driver. On the other hand, w-SHARP implements the clock synchronization using the beacon-based messaging and HW timestamps. As in the previous architecture, this one uses a common PHC for both IPs. Thus, the same CDC digital circuit used in the Ethernet TSN – IEEE 802.11 architecture has been implemented to translate the PHC counter from the Ethernet TSN clock domain to the wireless one.

The Ethernet TSN – w-SHARP domain translator runs two independent OS over the two ARM cores included in the SoC-FPGA platform. The first ARM core runs Free Real-Time Operating System (FreeRTOS) and the second one runs a Linux OS. On the one hand, the core running the FreeRTOS application provides the basic configuration for w-SHARP, generates the time-critical data, and runs a proprietary PTP-like daemon to perform the wireless clock synchronization. On the other hand, the Linux OS implements the connectivity with the Ethernet TSN network and runs a PTP daemon configured as a slave to synchronize the Ethernet TSN interface with the network.

The ARM cores exchange data through the Inter-Processor Communication (IPC) mechanism. The IPC is mainly used to access the w-SHARP modem through an IEEE 802.11 standard interface, which is used to send/receive best-effort frames from the Linux network stack. On the other side, FreeRTOS processes those frames and sends/receives the data from/to the w-SHARP modem. Linux has been also configured as a router between Ethernet and w-SHARP networks, providing full connectivity of best-effort data among the domains.

Finally, and in order to ensure seamless TSN operation among domains, the time-critical data generated by the wired-wireless TSN end devices have to be scheduled with minimum latency in the domain translator. This scheduling is implemented by the Ethernet TSN – w-SHARP bridge HW module. This implementation and the configuration of the TSN paths to minimize the E2E latency is out of the scope of this paper and it will be addressed in future works.

As in the case of the Ethernet TSN – IEEE 802.11 architecture, the end device (w-SHARP STA, Fig. 7) does not include the

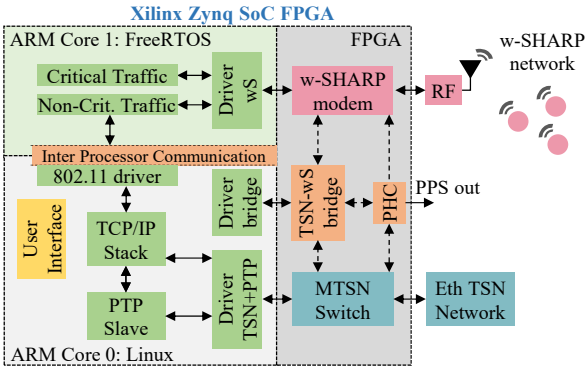

Fig. 6. Domain translator between Ethernet TSN and w-SHARP.

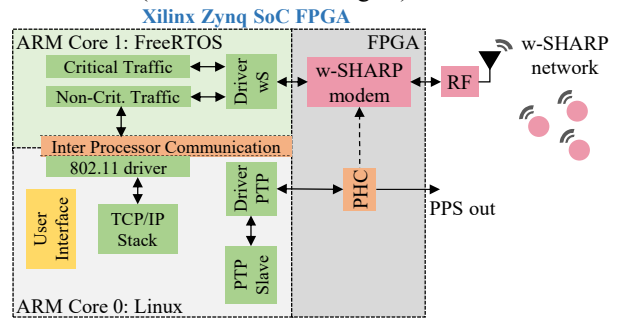

Fig. 7. w-SHARP end device in the Ethernet TSN – w-SHARP architecture.

Ethernet TSN interface. The w-SHARP device driver running over FreeRTOS synchronizes the PHC using the timestamps received from the beacons of the w-SHARP AP and the HW timestamps performed by the local NIC.

## V. ANALYSIS OF CLOCK SYNCHRONIZATION IN MULTI-HOP HYBRID NETWORKS

This section presents an analysis used to estimate the attainable clock synchronization in Hybrid TSN networks based on the error introduced in each of the network hops. First, the clock model is presented. Then, the clock synchronization error sources of the hybrid architectures are characterized. This analysis is used in the next sections to estimate the clock synchronization error.

### A. Clock models

Eq. (3) presents the clock model used in the analysis of the clock synchronization error

$$C(t) = t \cdot (1 + \rho(t)) + t_o, C[n] = C(t)|_{t=T_s \cdot (n+\theta)}, \quad (3)$$

being $C(t)$ the representation of the time in the continuous domain, $C[n]$ the discrete-equivalent time sampled at the PHC, $T_s$ the sampling period, $\theta$ the initial phase of the oscillator, $\rho(t)$ the phase drift of the oscillator, and $t_o$ the clock offset. A slave clock $C_S(t)$ is considered synchronized to the master clock $C_M(t)$ when their continuous version, i.e., $C_S(t)$ and $C_M(t)$ are equal. That is, when $\rho(t)$ and $t_o$ are equal in both clocks.

For the sake of simplicity, we have assumed that $\rho(t)$ is perfectly estimated and compensated by the network devices. Therefore, the analysis is mainly focused on the clock synchronization error introduced into $t_o$ by 1) the imprecisions in the timestamps and 2) in the CDC circuit to perform the PHC translation.

### B. Network synchronization hop

The translation of the timing from a master device to a slave device (network synchronization hop) introduces a clock synchronization error that strongly depends on the timestamping quality. For HW timestamping, these sources are: 1) the timestamping resolution bound (found for both wired and wireless), and 2) in the case of wireless links, the error introduced by the multipath and time-variant behavior of the wireless channel.

*1) Timestamping resolution bound*

The digital transceiver of either wired-wireless interfaces operate at a given frequency or clock period ($T_s$). The timestamping unit and PHC are typically built within the transceiver and are fed with the same clock source. As a result, the timestamps taken by the baseband processor are an integer multiple of $T_s$, introducing an error $\delta_r$ caused by the limited resolution. This error is modeled as a uniform distribution with bounds $[-T_S/2, T_S/2)$. Consequently, the minimum and maximum errors due to the resolution bound are

$$\min(|\delta_r|) = 0, \quad (4)$$

$$\max(|\delta_r|) = \frac{T_s}{2}. \quad (5)$$

For the sake of illustration, a Gigabit Ethernet interface operates at 125 MHz ($T_s = 8$ ns). Therefore its $\delta_r$ is bounded to $\pm 4$ ns. On the wireless side, the sampling period is typically a multiple of the inverse of the BW. For instance, the modems used in this work presents 20 MHz BW and $T_s = 50$ ns, i.e., $\delta_r$ for the wireless interfaces is bounded to $\pm 25$ ns. Note that, because of the internal structure of the wireless transceivers used in this work, only the Rx timestamps are affected by the timestamping resolution bound, whereas Ethernet PHYs have a resolution bound in both ingress and egress timestamps.

*2) Wireless channel delay*

In moving scenarios, wireless channels may have significant delay variations and can behave as asymmetric depending on their propagation properties (multipath and time-variation). Basically, a signal transmitted to the wireless medium is distorted by the channel multipath. The multipath generates several signal replicas that are received at different instants at the receiver. The preamble detector of the receiver tries to identify the first signal replica to take the timestamp. However, the preamble detectors are typically designed to find the replica with the highest power. Thus, depending on the multipath, it may detect secondary replicas instead of the first one. As a result, some timestamps may have a significant error because of the misdetection of the first component.

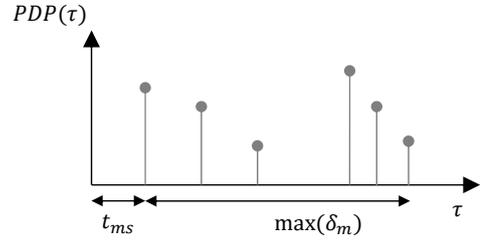

Fig. 8. Generic Power Delay Profile (PDP).

The error introduced by the multipath ($\delta_m$) depends on the specific propagation properties of the scenario and thus it cannot be modeled in a generic way. Nonetheless, an upper error bound can be provided based on the stochastic wireless channel properties $\max(\delta_m)$. Fig. 8 shows a generic Power Delay Profile of a wireless channel, which is the average of the channel impulse responses. $\max(\delta_m)$ equals the maximum difference between the delay of the first component of the channel and the last one. Therefore, a timestamp affected by the maximum possible error due to $\delta_m$ is expressed as follows

$$t'_2 = t_A + \max(\delta_m), \quad (6)$$

being $t_A$ the time of arrival of the first signal replica according to the PDP. The clock synchronization error introduced by $\delta_m$ depends on whether the messaging scheme is two-way or one-way. For two-way messaging, $\max(\delta_m)$ will be halved thanks to the channel delay calculations

$$\tilde{t}_o = t'_2 - t_1 + \frac{-t'_2 + t_1 - t_4 + t'_3}{2} = \frac{\delta_m}{2} + \frac{t_A + t_1 - t_4 + t'_3}{2}, \quad (7)$$

whereas $\max(\delta_m)$ is directly propagated to the $t_o$ estimation in the one-way messaging

$$\tilde{t}_o = t'_2 - t_1 = t_A - t_1 + \max(\delta_m). \quad (8)$$

$\max(\delta_m)$ can range from a few ns to more than 1 μs for indoor propagation upon the specific propagation conditions. Besides, note that the situation where the timestamping error is maximized due to $\delta_m$ is very unlikely due to the stochastic behavior of wireless channels. Finally, in the case of the one-way messaging, the bias in the clock synchronization due to the channel propagation delay ($t_{ms}$) is also an error source. $t_{ms}$

depends on the distance between the communicated nodes and for indoor propagation it can be assumed to be in a range of from a few ns to up to 100 ns (30 meters).

C. PHC translation

The PHC translation between two asynchronous clock domains can be modeled as a quantization error due to the sampling instant differences between the source domain PHC ($C_{Src}$) and the destination domain PHC ($C_{Dst}$) with clock periods of $T_{Src}$ and $T_{Dst}$ respectively. Let us assume that $T_{Src} \geq 4 \cdot T_{Dst}$ to be able to translate the time from the source to the destination without metastability issues using the CDC circuit shown in Fig. 5. The error in the PHC translation process ($\delta_{PHC}$) can be assumed to be the time difference between $C_{Src}$ and the time values sampled at the destination, as shown in Fig. 9. Note that $C_{Dst}$ samples include a calibration component of $T_{Src}/2$ to obtain a $\delta_{PHC}$ mean value equal to 0. Without loss of generality, we have assumed that $C_{Src}$ is our time reference so $\rho_{Src}(t) = 0$. Therefore, the error can be modeled using the sampling periods of each clock and $\rho_{Dst}$

$$\delta_{PHC}[n] = T_{Dst} \left( \rho_{Dst}[n] - \frac{T_{Src}}{2} + T_{Src} \left\lfloor \frac{1}{T_{Src}} \cdot (\rho_{Dst}[n] + T_{Dst} \cdot n) \right\rfloor \right), \quad (9)$$

being $\rho_{Dst}[n]$ the phase of $C_{Dst}$ at its sampling instants ($n$). Finally, the maximum and minimum $\delta_{PHC}[n]$ can be directly obtained from (9)

$$\min(|\delta_{PHC}[n]|) = 0, \quad (10)$$

$$\max(|\delta_{PHC}[n]|) = \frac{T_{Src}}{2}. \quad (11)$$

Eq. (9) shows that $\delta_{PHC}$ strongly depends on $\rho_{Dst}[n]$. If we assume that $\rho_{Dst}[n]$ slowly varies over $n$ in a random way, we can assume that, in the long term, $\delta_{PHC}$ will follow a uniform distribution with bounds $[-T_{Src}/2, T_{Src}/2)$. To sum up, this analysis highlights that the error in the PHC translation ($\delta_{PHC}[n]$) only depends on the $C_{Src}$ sampling period ($T_{Src}$) and that it is bounded to $\pm \frac{T_{Src}}{2}$.

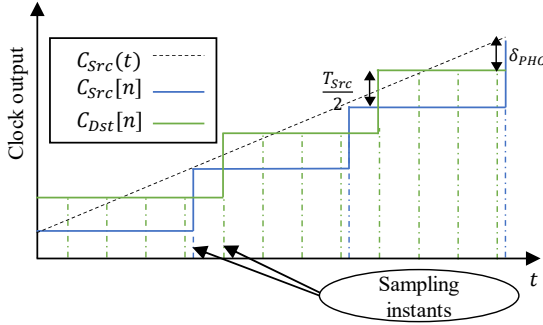

Fig. 9. PHC translation process.

VI. MEASUREMENTS SETUPS

This section presents the measurement setups used to evaluate the clock synchronization performance of the two Hybrid TSN architectures. For this purpose, three different setups with different objectives have been defined. The first setup involves the evaluation of the E2E clock synchronization performance of a hybrid TSN – 802.11 network under ideal conditions. In essence, the objective of this setup is to validate the proposed wired-wireless clock synchronization translation mechanisms and architecture and measure its minimum performance bound. The second setup is used to evaluate the wireless clock synchronization under known and repeatable wireless conditions using a channel emulator for both Ethernet TSN – 802.11 and Ethernet TSN – w-SHARP. This setup enables the evaluation of the solutions over a wide range of wireless conditions without leaving the laboratory. The last setup is used to evaluate the clock synchronization performance transmitting over the air. This setup also covers both Ethernet TSN – 802.11 and Ethernet TSN – w-SHARP architectures.

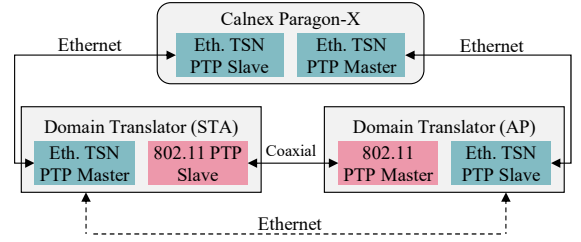

Fig. 10. Block diagram of the Measurement setup based on Calnex Paragon-X.

A. Ethernet TSN – 802.11 measurement setup using the Calnex Paragon-X

This setup uses the Calnex Paragon-X equipment [30] to validate and test the Ethernet TSN – 802.11 architecture under controlled conditions (see Fig. 10). The Calnex Paragon X provides the emulation of a PTP master (GMC) and a PTP slave and it is able to measure the clock synchronization error between its ports allowing the testing of a PTP chain. This setup uses two Ethernet TSN-802.11 domain translators. The 802.11 modem of the first clock domain translator is configured as AP and the 802.11 modem of the second clock domain translator is configured as STA. Each clock domain translator runs two PTP daemons which are configured as slave or master upon the task of the interface, working as a boundary clock. In the case of the domain translator AP, the Ethernet TSN interface is configured as a PTP slave and the 802.11 interface is a PTP master. On the domain translator STA side, the 802.11 interface is a PTP slave and the Ethernet TSN interface is a PTP master. Since the objective of this setup is to evaluate the minimum attainable clock synchronization over controlled conditions, the wireless interfaces have been connected using coaxial wires.

PTP in the wired segment has been configured using the default configuration values of the Linux PTP daemon. Namely, the configuration has been set to one-step and multicast addressing, the messaging exchange period has been set to 1 s, and the time filtering method has been set to PI with constants $K_p = 0.7$ and $K_i = 0.3$. On the wireless side, PTP has been set to one-step and unicast addressing. The filtering method has been also set to PI with the same constants, and the Sync period has been reduced from 1 second to 1/8 seconds to compensate for the lower timestamping resolution provided by 802.11.

Regarding the expected performance, this network includes 2 Ethernet TSN hops, two PHC translation hops, and one wireless hop. Therefore, the maximum expected error using oscillators without frequency drift can be expressed as follows

$$\max(\delta) = 4 \cdot \max(\delta_{r,Eth}) + 2 \cdot \max(\delta_{PHC}) + \max(\delta_{r,w}), \quad (12)$$

being $\delta_{r,Eth}$ the timestamping resolution bound of Ethernet ($\pm 4\,ns$) and $\delta_{r,w}$ the timestamping resolution bound of the wireless modems ($\pm 25\,ns$). Note that $\delta_{r,Eth}$ is multiplied by 4 because this network topology includes two Ethernet links and because both ingress and egress Ethernet timestamps are affected by the timestamping resolution bound $\delta_{r,Eth}$.

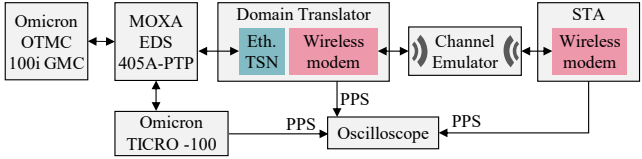

Fig. 11. Block diagram of setup used to evaluate the architectures using the wireless channel emulator.

Finally, and for the sake of comparison between the synchronization of only-Ethernet TSN and Ethernet TSN – 802.11, we have considered an alternative setup where the domain translators are directly connected through Ethernet instead of with wireless, building a three-hop Ethernet network, as depicted in Fig. 10.

### B. Measurement setup using a wireless channel emulator

This setup has been used to evaluate the attainable clock synchronization of both architectures over a set of different wireless propagation conditions, from office to industrial. Fig. 11 displays the block diagram and a photo of the setup. An Omicron GPS GMC drives the time source, which is distributed through a Moxa EDS-405A-PTP Ethernet switch to the domain translator with the wireless interface (either w-SHARP or 802.11). The wireless interface then extends the synchronization to the STA. Since the GPS GMC does not have a PPS output, we have also connected to the switch an Omicron TICRO-100 PTP time converter. The TICRO-100 PTP has a PPS which has been considered as the network base time. The PPS output of the TICRO-100 PTP has been used to validate the synchronization of the Ethernet TSN segment.

The clock synchronization is evaluated by the time difference between the rising edge of the PPS of each device. Specifically, the PPS signals of the Omicron TICRO-100 and the domain translator have been used to measure the synchronization of the Ethernet TSN segment and the PPS of the domain translator and STA have been used to evaluate the wireless synchronization.

To ensure repeatable wireless conditions, the measurements have been carried out using a channel emulator. The channel emulator used in the experiments has been tuned to operate in the 2.4 GHz band with a bandwidth of 100 MHz. It has four bidirectional ports with a delay from port to port of 1135 ns, and a minimum fixed attenuation of 18 dB. The emulator can hold a wireless channel model up to 10 taps that can be programmed with different gains and fading profiles. It supports two fading distributions: Rayleigh and Rice; and three Doppler spectrum shapes: Jakes, Bell, and Gaussian [31].

TABLE II. WIRELESS CHANNEL MODELS.

|  | Scenario | rms Delay Spread[ns] | max($\delta_m$) [ns] | Fading Model |
|---|---|---|---|---|
| WLAN A | Small Office | 50 | 390 | Rayleigh+Jakes |
| WLAN C | Large Office | 150 | 1050 | Rayleigh+Jakes |
| IWLAN A | Industrial | 29 | 140 | Rayleigh+Jakes |
| IWLAN B | Industrial | 89 | 600 | Rayleigh+Jakes |

We have carried out the clock synchronization measurements over four wireless channels (Table II). The first two channels (Wireless Local Area Network (WLAN) A and C [32]) correspond with a small and large office space respectively. The third and fourth channels, named as IWLAN A and IWLAN B (Scenario 7 and CM8 in [33]), represent two industrial scenarios with low and medium dispersion. Besides, we have used two variation speeds (10 and 30 km/h) which are typical speeds of mobile robotics scenarios.

Concerning the clock synchronization, the 802.11 modem is configured with the PTP default configuration values and with a frame exchange periodicity of 1/8 s. In the case of w-SHARP, its clock synchronization is performed through the beacon-based messaging with $T_{Sync} = 500$ μs, and with a PI filter with constants $K_p$ and $K_i$ of 0.1 and 0.01 respectively. These values have been selected based on empirical measurements over non-dispersive channel conditions. It must be noted that, since w-SHARP is based on the beacon-based synchronization scheme, it cannot estimate the channel delay. Therefore, the channel emulator base delay (1135 ns) has been pre-calibrated in the clock synchronization calculations.

Regarding the expected performance, this network includes 2 Ethernet TSN hops, one PHC translation hop, and one wireless hop. The expected synchronization depends on whether the hop is performed with 802.11 or with w-SHARP. The maximum expected error for oscillators without frequency drift for Ethernet TSN - 802.11 is as follows

$$\max(\delta) = 4 \cdot \max(\delta_{r,Eth}) + \max(\delta_{PHC}) + \max(\delta_{r,w}) + \frac{\max(\delta_m)}{2}, \quad (13)$$

whereas the maximum expected error for Ethernet TSN - w-SHARP would be

$$\max(\delta) = 4 \cdot \max(\delta_{r,Eth}) + \max(\delta_{PHC}) + \max(\delta_{r,w}) + \max(\delta_m) + \max(t_{ms}). \quad (14)$$

### C. Measurement setup transmitting over the air

Finally, and in order to evaluate the clock synchronization performance of both wireless architectures in a more realistic setup, we have performed a set of experiments in a large office environment with a size of 19 x 25 meters, similar to a medium-sized industrial facility [34]. These experiments have been done over the 2.4 GHz band with no 802.11 networks and devices operating in the same or adjacent channels. This free spectrum condition is recommended to focus the experiment on the wireless propagation characteristics and its synchronization results due to the cannel.

Taking into account the constraint of the length of the oscilloscope's probes, it is not possible to simultaneously measure the PPS of devices placed at a long distance. Therefore, we have measured the relative synchronization error between the STA and the OMICRON TICRO-100, as shown in Fig. 12. The latest has been placed near the STA, and it has been synchronized through Ethernet with a large 10-meter Ethernet wire. Two different positions have been considered. In the first position, the devices were placed in the same table at 0.5 meters and with direct line of sight. In the second position, the devices were placed at 10 meters and without direct line of sight.

This setup uses the same PTP configuration parameters described in subsection V.B.

## VII. RESULTS

The results have been divided into three subsections following the structure of section VI. The first subsection shows the E2E synchronization results for the Ethernet TSN – 802.11 architecture using the Calnex Paragon X. The second subsection displays the effects of the wireless channel in the wireless clock synchronization. Finally, the last subsection shows the performance of both architectures with the wireless links carried out over the air. The experiments have been run for a total of

1000 seconds and 1000 samples have been taken (1 sample per PPS rising edge).

### A. Ethernet TSN – 802.11 E2E measurement results

TABLE III. CLOCK SYNCHRONIZATION MEAN AND STD. DEVIATION FOR THE ETHERNET TSN – 802.11 SETUP BASED ON CALNEX MEASUREMENT.

|  | μ [ns] | σ [ns] | \|μ\| + 3σ [ns] | max (δ) [ns] |
|---|---|---|---|---|
| **Ethernet (3 hops)** | -4.3 | 18.7 | 60.4 | 24 |
| **802.11 (1 hop), Ethernet (2 hops)** | -5.5 | 16.6 | 55.3 | 73 |

The clock synchronization results of the Ethernet TSN – 802.11 architecture using the Calnex Paragon X instrument along with the theoretically expected values are reported in Table III. As can be seen, the clock synchronization error of the Hybrid TSN – 802.11 architecture presents an average (μ) of -5.5 ns and an std. deviation (σ) of 16.6 ns. The results for the Ethernet TSN – 802.11 architecture are in the range of what is expected based on the theoretical analysis, assuming that |μ|+3σ equals the maximum deviation. The reported synchronization performance is in the range of the required performance for TSN operation [9]. Besides, and for the sake of comparison, we also performed the measurements for a three-hop Ethernet TSN network. As can be seen, the performance with either the wired or the wireless link is very similar. Nonetheless, In the case of the 3-hops Ethernet TSN network, the predicted performance is much better than the measured error. This behavior can be explained from an implementation point of view, where MTSN IP would require optimization to reach the theoretical bound. The synchronization provided by the MTSN is enough for networks with a small number of hops, though further optimizations may be required in large-scale industrial networks.

### B. Measurement results over the wireless channel emulator

*1) Ethernet TSN – 802.11 results*

The clock synchronization performance for the 802.11 link over the different wireless channels is summarized in Table IV for the considered channel models and for an ideal Additive White Gaussian Noise (AWGN) channel. As can be seen, the best results are obtained for the channels with the lowest energy dispersion. For instance, in the AWGN channel, the results show a standard deviation of 10.8 ns and a mean value of 5.5 ns.

It can be counterintuitive that industrial channel models present better results because they are typically harsher than other conditions. However, the difference in the results is totally explained by the different rms delay spreads of each wireless channel model. Specifically, the measurement shows that σ grows with the channel rms delay spread, from 10.8 ns in the AWGN channel to 20.2 ns for WLAN C at 30 km/h and μ also grows from 4.9 ns to 19.1 ns. Both effects are explained by means of the effects described in subsection V.B.2). Under multipath conditions, the receiver may detect different secondary signal replicas instead of the first one, which introduces a jitter in successive timestamps. In addition to that, small differences in the calibration of the receivers can introduce a bias in the probability of detection of the first or secondary components. Basically, a receiver prone to detect more secondary components will introduce a delay asymmetry. The delay asymmetry is then translated to a bias in the synchronization (μ).

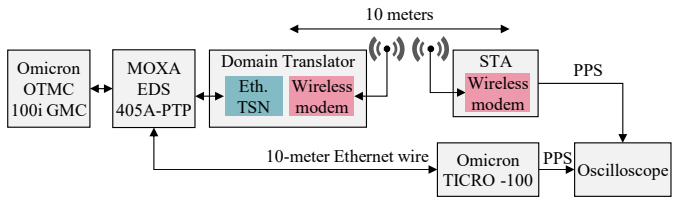

Fig. 12. Setup used to evaluate the hybrid architectures over the air.

Regarding the results as a function of the channel variation speed, it can be seen that it has little influence on the resulted synchronization. This effect is explained by means of the relation between of the channel coherence time $T_c$ and the PTP messaging exchange duration. Basically, the channel coherence time $T_c$ is larger than the time taken to perform the PTP exchange for both variation speeds, thus every frame of a specific PTP frame exchange is affected by a similar wireless propagation.

TABLE IV. CLOCK SYNCHRONIZATION PERFORMANCE OF 802.11 FOR DIFFERENT WIRELESS CONDITIONS USING CHANNEL EMULATOR.

|  | 0 km/h | | 10 km/h | | 30 km/h | | max ($\delta_w$) [ns] |
|---|---|---|---|---|---|---|---|
|  | μ [ns] | σ [ns] | μ [ns] | σ [ns] | μ [ns] | σ [ns] |  |
| **AWGN** | 4.9 | 10.8 | | | | | 25 |
| **WLAN A** | | | 14.6 | 13.1 | 15.2 | 13.4 | 220 |
| **WLAN C** | | | 18.0 | 19.7 | 19.1 | 20.2 | 550 |
| **IWLAN A** | | | 7.2 | 11.3 | 8.2 | 11.9 | 95 |
| **IWLAN B** | | | 9.3 | 17.2 | 7.8 | 19.7 | 325 |

Regarding the maximum theoretical bound, it can be seen that the maximum expected error in the wireless link

$$\max(\delta_w) = \max(\delta_{r,w}) + \max(\delta_m) \qquad (15)$$

is dominated by the asymmetry introduced by the multipath propagation $\delta_m$, which depends on the specific wireless channel propagation. Specifically, $\max(\delta_w)$ equals 95 ns for the wireless channel with the lowest rms delay spread, whereas $\max(\delta_w) = 550$ ns is expected for WLAN C. The situation where $\delta_w$ is maximized has very low occurrence probability and thus some techniques could be used to reduce the total error. For instance, a timestamping filter could be used to detect timestamps that are very far from the expected timestamp value [35]. Finally, the maximum expected error of the hybrid network max (δ) equals 143 ns and 598 ns for the IWLAN A and WLAN C respectively. This result highlights that the highest error source is the still unpredictable behavior of the channel.

*2) Ethernet TSN – w-SHARP results*

This subsection presents the clock synchronization results of the w-SHARP link over different wireless conditions using the wireless channel emulator. The performance results are summarized in Table V and described as follows.

In the case of the AWGN channel, the results show a clock synchronization performance with a very low standard deviation of 1.9 ns and a mean value of -5 ns. Note that the result is quite different from the theoretical one (25 ns) because we are using a low messaging exchange periodicity (500 μs) and a narrow PI filter to average the timestamping error, which are not accounted in the theoretical analysis. On the other hand, for wireless channels with multipath, and as opposed to 802.11, w-SHARP uses the one-way messaging and it is not able to estimate the channel delay. As a consequence, μ presents a large bias. In addition to that, σ also grows because of the multipath propagation. Compared with the results over 802.11, the first impression is that w-SHARP is affected by the movement speed of the nodes, though, surprisingly, the results are better for larger speeds than for lower speeds. These results are explained by the low periodicity of the synchronization frames and the PI filter tuning. Basically, if $T_c \gg T_{Sync}$, w-SHARP synchronization will follow the channel delay variations caused by the multipath. Consequently, the w-SHARP STAs are following the higher energy path of the channel, and thus the timestamps $t'_2$ are suffering strong variations. On the contrary for high movement speeds, each synchronization frame is affected by different uncorrelated realizations of the wireless channels and thus the w-SHARP synchronization stays around the average channel delay, whose variation follows a random process. As a result, the w-SHARP STAs are not able to follow the timestamps $t'_2$ variation due to channel delay variation and stay around the mean value.

TABLE V. CLOCK SYNCHRONIZATION PERFORMANCE OF W-SHARP FOR DIFFERENT WIRELESS CONDITIONS USING THE CHANNEL EMULATOR.

|  | 0 km/h | | 10 km/h | | 30 km/h | | max ($\delta_w$) [ns] |
|---|---|---|---|---|---|---|---|
|  | μ [ns] | σ [ns] | μ [ns] | σ [ns] | μ [ns] | σ [ns] |  |
| AWGN | -5.19 | 1.9 |  |  |  |  | 25 |
| WLAN A |  |  | 47.3 | 26.7 | 38.4 | 20.2 | 415 |
| WLAN C |  |  | 91.6 | 53.2 | 77.1 | 20.3 | 1075 |
| IWLAN A |  |  | 34.5 | 15.2 | 16.8 | 14.9 | 165 |
| IWLAN B |  |  | 45.1 | 32.9 | 53.9 | 20.5 | 625 |

Finally, regarding the performance results as a function of the rms delay spread, the experiments with high time-dispersive channels present a larger σ. The difference in σ among channels is more significant in low-mobility configurations, because of the aforementioned reasons. Nonetheless, for higher mobility configurations, there are no significant differences between the different wireless channels. Even so, the synchronization can be enough for most Wireless TSN configurations, especially if an appropriate channel delay calibration is performed in the commissioning of the network devices.

As in the theoretical results for Ethernet TSN – IEEE 802.11, max($\delta$) is dominated by wireless propagation conditions. Because w-SHARP uses the one-way messaging, both $\delta_m$ and $t_{ms}$ influence the maximum expected clock synchronization error. Specifically, max($\delta_w$) equals 165 ns for IWLAN A and 625 ns for IWLAN C for a coverage range of 30 meters ($t_{ms} = 100\ ns$). In this case, the timestamping filtering may not be as evident as in the two-way messaging, because it may not be clear whether a variation from the expected timestamp value comes from a variation in $t_{ms}$ or from the multipath $\delta_m$. Therefore, some specific works may be carried out to evaluate if these errors can be distinguished or not.

### C. Measurement results transmitting over the air

This subsection presents the results of the measurements carried out over the air in a large open office environment. The results are summarized in Table VI and explained as follows. We have not estimated max ($\delta_w$) in this case because the wireless channel propagation has not been characterized. The results from these experiments are in line with our findings in the experiments using the channel emulator. Regarding 802.11, there are small differences between the results at 0.5 meters or 10 meters. Basically, the 802.11 link can compensate for the channel delay thus the distance between the nodes does not influence the μ. Also, there is virtually no difference in the σ because the experiment was run over static wireless conditions.

TABLE VI. SYNCHRONIZATION PERFORMANCE TRANSMITTING OVER THE AIR.

|  | Ethernet TSN - 802.11 | | Ethernet TSN - w-SHARP | |
|---|---|---|---|---|
| Experiment | μ [ns] | σ [ns] | μ [ns] | σ [ns] |
| 0.5 meter | 7.2 | 12.4 | 10.1 | 16.3 |
| 10 meters | 7.8 | 13.1 | 73.3 | 19.4 |

The results are more dispersed in the case of w-SHARP because the one-way messaging cannot compensate for the channel delay. In this case, it can be seen a μ difference of 63.2 ns between the experiments. A difference of 63.2 ns is in line with what is expected for the one-way messaging and the setups. First, the difference in the propagation delay for both experiments in free conditions would be around 30 ns based on the distance between the nodes in both experiments. In addition to that, the second experiment was run without a direct line of sight and so the propagation is performed through reflections of the environment, which further increases the propagation delay. Finally, regarding σ, there is a slight deterioration when the distance between the nodes is increased. The deterioration mainly comes from the increased multipath in the experiment at 10 meters.

### D. Results summary

Table VII summarizes the attainable clock synchronization performance under different wireless channel impairments. The table highlights three parameters that define the wireless channel models: movement or coherence time $T_c$, propagation delay $t_{ms}$ and rms delay spread. On the other side, it summarizes the influence of such parameters over the synchronization performance in terms of delay average and standard deviation.

As demonstrated by the results attained in each experiment, there is a direct relation between the channel's delay spread and the clock synchronization results for both w-SHARP and 802.11, where the standard deviation grows with the channel delay spread (see subsection V.B.2). Besides this, $t_{ms}$ has a direct influence in the synchronization μ in the case of w-SHARP because its messaging scheme does not include channel delay compensation, whereas it has no influence in 802.11.

Concerning the dynamics of the wireless channel, the achieved performance depends on the synchronization messaging protocol. On the 802.11 side, PTP is able to calculate the channel delay by means of the message exchange and therefore the channel variation speed does not have a great influence in the synchronization. On the other hand, w-SHARP has worse synchronization results for low movement speed than for high movement speeds because of the relation between $T_c$ and $T_{Sync}$.

TABLE VII. SYNCHRONIZATION PERFORMANCE SUMMARY.

|  | 802.11 PTP | | w-SHARP | |
|---|---|---|---|---|
| Perturbation | Average (μ) | Std Dev (σ) | Average (μ) | Std Dev (σ) |
| ↑ Movement Speed (↓ Coherence Time $T_c$) | No Change | No Change | Decrement | Decrement |
| ↑ rms Delay Spread | No Change | Increment | No Change | Increment |
| ↑ Delay $t_{ms}$ | No Change | No Change | Increment | No Change |

To sum up, the experiments and results described in this section demonstrate that the channel propagation conditions influence the synchronization, though the attained performance is enough to fulfill the required performance for Wireless TSN operation even for the most challenging wireless propagation considered.

## VIII. CONCLUSIONS

This paper dealt with the implementation and evaluation of high-performance clock synchronization over Hybrid TSN (wired-wireless TSN) networks. The paper started with a review of the state-of-the-art and the different techniques commonly used to perform clock synchronization over wireless. Afterward, we presented two hybrid network architectures with E2E clock synchronization capabilities, one based on Ethernet TSN – 802.11, and the second one based on Ethernet TSN – w-SHARP. Using these architectures, we evaluated the attainable clock synchronization. The experiments showed that the developed architectures are well enough to fulfill the TSN requirements in terms of clock synchronization.

Nevertheless, the road to hybrid TSN still presents numerous obstacles, most of them in the wireless domain. In the first place, several other TSN features apart from synchronization must be supported by the wireless system to be considered "Wireless TSN", e.g. frame scheduling, path control and reservation, frame preemption, etc. In the second place, the data translation and routing between the wired and wireless domains must be done in a fast and deterministic way to synchronize the TSN schedulers in each node and maintain the TSN flows over the wired-wireless links. Therefore, significant design and implementation efforts are required in the wireless domain and in the domain translation to ensure appropriate Hybrid TSN operation.

As a future line of work, we will study the complete integration of w-SHARP and Ethernet TSN to demonstrate the suitability of Hybrid TSN in industrial applications. The integration will include the synchronization (already demonstrated in this work), and the scheduling between the Ethernet and the wireless TSN domains timeliness capabilities expected from a TSN communication system.